\documentclass[aps,twocolumn,showpacs,nofootinbib]{revtex4-1}
\usepackage{graphicx}   
\usepackage{latexsym}   
\usepackage{enumerate}
\usepackage{color}
\usepackage{ulem}
\usepackage{float}

\begin{document}

\title{Effects of fluctuations and color-neutrality in a finite volume}

\author{Christian Spieles$^1$, Marcus Bleicher$^2$ and Carsten Greiner$^2$}

\affiliation{$^1$~Frankfurt Institute for Advanced Studies, Ruth-Moufang-Strasse 1, D-60438 Frankfurt am Main, Germany\\
             $^2$~Institut f\"ur Theoretische Physik, Goethe-Universit\"at, Max-von-Laue-Strasse 1, D-60438 Frankfurt am Main, Germany}
             

\begin{abstract}
We investigate properties of strongly interacting matter in a schematic
model, based on the combined degrees of freedom of a non-interacting hadronic
phase and a non-interacting deconfined phase. It is found that in a finite system both phases contribute to the thermodynamic state due to fluctuations 
and that signatures of critical behviour like the divergence of statistical
quantities are damped. The constraint of color-neutrality leads to a volume-dependent shift of the
effective critical temperature, which follows a scaling law, independent of
the baryochemical potential. According to the model, observable baryon-number susceptibilities at a given $T$ and $\mu_B$ strongly depend on the system
size. Finally, we compare hadronization conditions from the model with hadrochemical fits to
experimental collider data, where a qualitatively similar system size dependence is extracted.
\end{abstract}

\maketitle

\section[]{Introduction}
The phase-structure of strongly interacting matter, described by Quantum
Chromodynamics, has long been a particular focus of theoretical and experimental
research (see Ref.~\cite{MeyerOrtmanns1996} for a review of basic
concepts). Within the last two decades, tremendous progress has been made and is
still under way, see, e.~g., Refs.~\cite{Bazavov2017,Guenther2018} and
Ref.~\cite{Luo2017}, respectively. However, it is found that two main
difficulties still impede a comprehensive theoretical understanding of the
whole phase-structure backed by experimental evidence:
One is the fact that rigorous calculations solving QCD for the baryon-rich
regime are still not feasable. The other challenge is posed by the limited spatial and temporal scales of
any experiment probing strongly-interacting matter under extreme conditions. Furthermore, local fluctuations of energy density within single heavy-ion collisions
may play a decisive role for the signatures under investigation (see \cite{Bleicher1998}.)
In the following, we revisit a schematic model of strongly
interacting matter \citep{Spieles1998,Spieles2019}, which allows to explore some possibly relevant features of {\it finite}
matter, even at  $\mu_B \gg 0$. We also report some new findings in
Secs.~\ref{sec:voldep} and \ref{sec:hadr}.

\section[]{The model}\label{sec:model}
The schematic two-phase model of strongly interacting matter presumes 
coexistence of two microscopically uncorrelated phases which are
connected only via the macroscopic configuration, namely
the volume fraction $\xi$ of one of the two phases (we choose the hadronic
phase, i.~e., $V_h=\xi V$ and $V_q=(1-\xi)V$).
In such a simplified set-up, any macroscopic configuration $\xi$
contributes with a probability  $p(\xi) \sim \exp [-\Phi(\xi)/T]$ to
the total system, $\Phi(\xi)$ being the grand canonical potential of the system
for this particular configuration \citep{Landau1976}. 

Since the partition
function of the total system factorizes into the partition fuctions of the
two individual phases for any fixed $\xi$, the grand canonical potential $\Phi$
of the total system in configuration $\xi$ can be expressed as
\begin{eqnarray}\label{eq:phixi}
\Phi_\xi(T,\mu_B,V)&=& [\varphi_h(T,\mu_B,\xi V)\xi  \\*
&+& \varphi_q(T,\mu_B,(1-\xi)V) (1-\xi)]V \nonumber
\; ,
\end{eqnarray}
where $\varphi_h$ and $\varphi_q$ are the densities of the grand canonical potential of the
hadron gas und the quark-gluon phase, respectively.
Any intensive thermodynamic quantity  $A(T,\mu_B,V)$ describing the total system is then given as
an expectation value according to the weight of  all possible
configurations:
\begin{eqnarray}\label{eq:model}
A(T,\mu_B,V)=\int_0^1 p(\xi;T,\mu_B,V)[A_h(T,\mu_B,\xi V)\xi  \\* 
 + A_q (T,\mu_B,(1-\xi)V)(1-\xi)] d\xi \, . \nonumber
\end{eqnarray} 
Note that for a system of infinite volume, the schematic model renders the Gibbs
equilibrium condition according to which the two phases only coexist at $T_C$, where
the pressures of the individual phases coincide, $p_h=p_q$.
The phase transition in this case is of first order.\footnote{For
sufficiently high values of the baryochemical potential this may well be in
accordance with the true properties of strongly interacting matter. At $\mu_B=0$,
however, lattice QCD does not exhibit a first order phase transition.}
For finite volumes, in contrast, the model equation (\ref{eq:model})
necessarily implies a smooth crossover of thermodynamic quantities which is
not characteristic of a first order phase transition:
However weak, fluctuations must lead to the presence of {\it both} phases for any value of $T$ and
$\mu_B$, i.~e. $0< \langle \xi \rangle < 1$. 

The model equation of state of the hadronic phase is based on an ideal
relativistic quantum gas of well established non-strange baryon and meson resonances up to masses of
2~GeV.  Its density of the grand canonical potential is
\begin{equation}
\varphi_h = -\sum_i \frac{g_i}{6 \pi^2} \int_0^{\infty} \frac{dp}{E_i} \frac{p^4}{\exp [(E_i-\mu_i)/T] \pm 1}  \quad  ,
\end{equation}
where "$+$" stands for fermions and "$-$" for bosons, $g_i$ denotes the degeneracy of particle species $i$. 
$E_i=\sqrt{p^2+m_i^2}$ is the energy of particle species $i$ and $\mu_i$ its chemical potential. 
All thermodynamic quantities are then corrected by the Hagedorn factor $1/(1+\epsilon/4B)$ \citep{Hagedorn1980}, 
where $\epsilon$ is the ideal gas energy density and $B$ is the bag pressure. 
The deconfined phase is thought of as an ideal relativistic quantum gas of
massless quarks and gluons in a cavity, held together by the bag pressure. 
In the case of two quark flavors and with the constraint of 
color-neutrality and fixed total momentum, the
corresponding density of the grand canonical potential for a spherical droplet
of
volume $V$ can be approximated according to \cite{Elze1986}: 
\begin{equation}\label{eq:partitionfunc}
\varphi_q =-T/V \left[ \ln{(\frac{1}{2}\sqrt{\frac{1}{3}\pi}\; C^{-4}
D^{-3/2})} + X - Y \right] + B \quad ,
\end{equation}
where 
\begin{eqnarray}
X  &=& \pi^2 VT^3 \times  [ \frac{37}{90} +
( \frac{\mu_q}{\pi T} )^2 + \frac{1}{2} (\frac{\mu_q}{\pi T})^4 ]  \nonumber \\* 
Y &=&  \pi (\frac{3V}{4\pi})^{1/3} T [\frac{38}{9} +2 (\frac{\mu_q}{\pi T})^2 ] 
\quad .
\end{eqnarray}
Note, that $Z_0=\exp{(X - Y)}$ is the grand partition function without overall
constraints. The remaining parameters for the color and momentum constraints in
(\ref{eq:partitionfunc}) are
\begin{equation}
C = 2 VT^3 [\frac{4}{3}+(\frac{\mu_q}{\pi T})^2 ] +
\frac{20}{3\pi}(\frac{3V}{4\pi})^{1/3}T   \quad 
\end{equation}  
and
\begin{equation}
D = 2 X - \frac{1}{3}Y \quad . 
\end{equation}
The bag constant is set to $B^{1/4}=215~{\rm MeV}$. For infinite volumes, this
corresponds to a critical temperature of $T^{\infty}_C \approx 155~{\rm MeV}$ at $\mu_B=0$
in the two-phase coexistence model.\footnote{This value agrees with the the chiral
transition temperature derived from lattice calculations (see Ref.~\cite{Bazavov2017} and
references within).}

\section[]{Volume-dependence of the effective critical temperature}\label{sec:voldep}
\begin{figure}[t]
\hspace*{0.3cm}
 \centerline{\includegraphics[width=320pt]{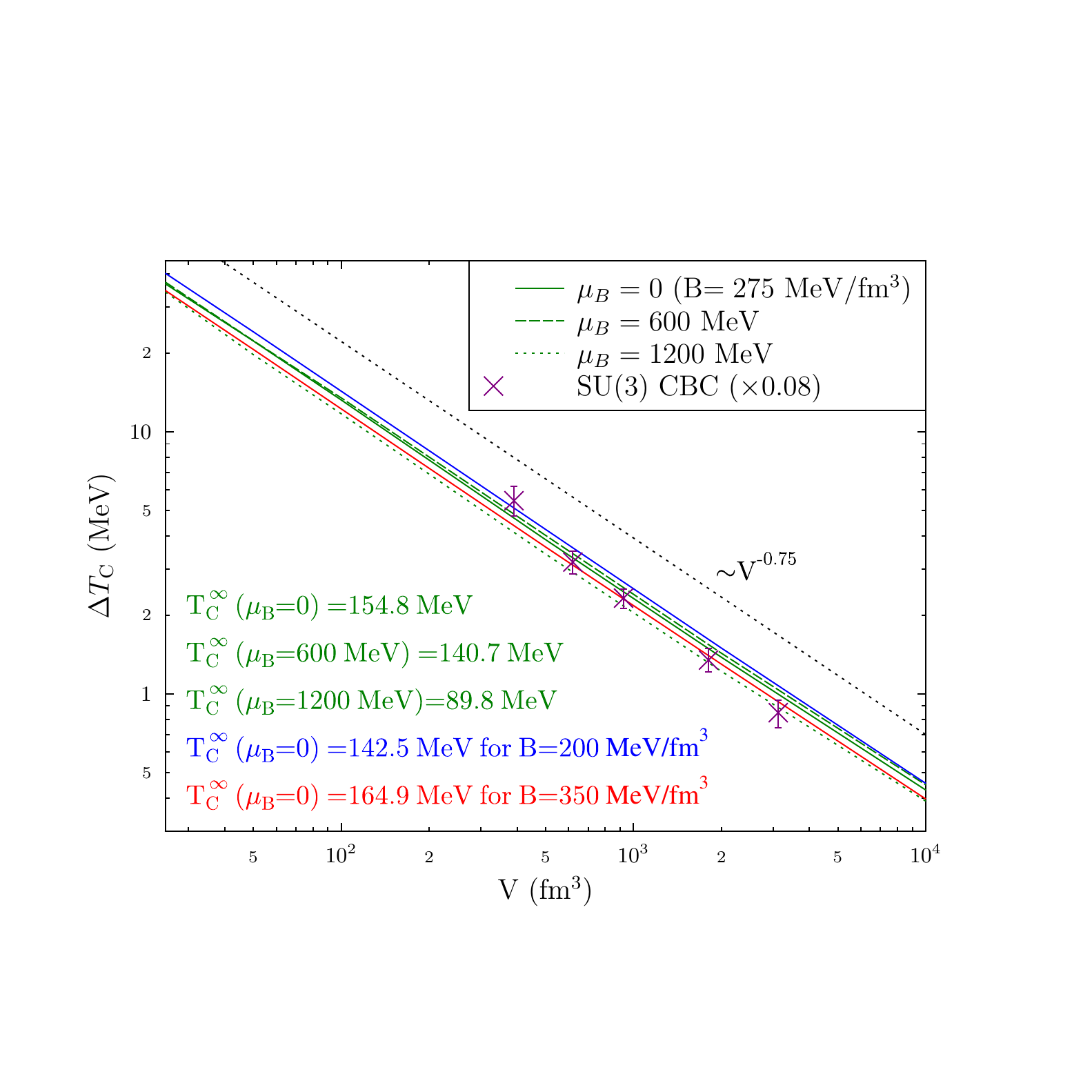} 
 }
\vspace{-2.cm}
 \caption{The shift of the effective critical temperature as a function of system volume according to the schematic two-phase model. 
Calculations for different values of baryochemical potential and different values of the bag pressure $B$ are shown
as colored lines. Also shown is the varation of the pseudo-transition temperature as compared to the infinite volume limit
 as derived from pure ${\rm SU(3)}$ lattice gauge theory with cold boundary conditions (CBC)
\citep{Berg2013}. Absolute values are scaled as indicated.}
\label{fig:deltatc}
\end{figure}

As was shown in Ref.~\cite{Spieles1998}, the color-singlet constraint
imposed on the model equation-of-state leads to a volume dependent shift of
the "effective critical temperature", $\Delta T_C (V) = T_C^{\rm
eff}(V)-T_C^{\infty}$. In the present study, we define this temperature as the point where
both phases contribute --- on the average --- with equal probability to the macroscopic
state of the system, i.~e. $\langle \xi (T_C^{\rm
eff}(V)) \rangle = 1/2$. The reason for the shift is that the quark-gluon
phase in a finite volume is associated with a smaller grand canonical potential density
than in the infinite volume limit. The effective number of degrees of
freedom in the quark-gluon phase is reduced in small systems, since only
color-neutral combinations of microstates are taken into account in the
partition function. 

It is expected that the volume dependence of the shift of the critical
temperature exhibits a universal scaling behaviour $\Delta T_C (V) \sim V^{-\lambda}$, which characterizes fundamental
properties of the physical system. E.~g., \cite{Binder1984} have advocated 
a value of $\lambda \approx 1$ as a signature of a first order phase
transition. Fig.~\ref{fig:deltatc} shows an analysis of $\Delta T_C(V)$
for different values of the baryochemical potential according to the
schematic model. In fact, although the respective scenarios reflect very different
physical conditions, a universal scaling exponent of $\lambda \approx 0.75$ can be
extracted from the numerical analysis. Note that not only the values of the
scaling exponent coincide but that also the absolute temperature shifts are
very close. Furthermore, the scaling law also holds for different values of
the bag constant $B$ --- the one system parameter that physically governs the
interdependence of the two phases, which are microsopically uncorrelated.

We contrast our result with findings of \cite{Berg2013} where the
deconfining phase transition of pure SU(3) lattice gauge theory has been
investigated by Markov-chain Monte Carlo simulations. In their study, the authors have
calculated small volume corrections for the pseudo-transition temperature in
a scenario of cold boundary conditions. We show the resulting temperature
shifts in Fig.~\ref{fig:deltatc} in order to compare it with the
scaling law of our model. 
As can be seen, the absolute temperature shift in the pure SU(3) lattice
gauge simulation for a given system size differs 
considerably from the schematic two-phase model. However, the volume dependence
indicates a scaling exponent in the range of $0.75 < \lambda < 1$, which is
compatible with the model result.
In this context, we would like to draw the attenation to the work of
\cite{Ladrem2005} where a similar schematic two-phase coexistence model
than the one presented in Ref.~\cite{Spieles1998} has been the basis for an analysis of finite-size
effects and scaling exponents in the deconfinement phase transition (at
vanishing net-baryon density). In their approach, the authors extract a temperature shift scaling exponent of $\lambda = 0.876
\pm 0.041$.

\section{Hadronization of a finite quark-gluon plasma}\label{sec:hadr}
\begin{figure}[tb]
\hspace*{0.2cm}
\centerline{\includegraphics[width=300pt]{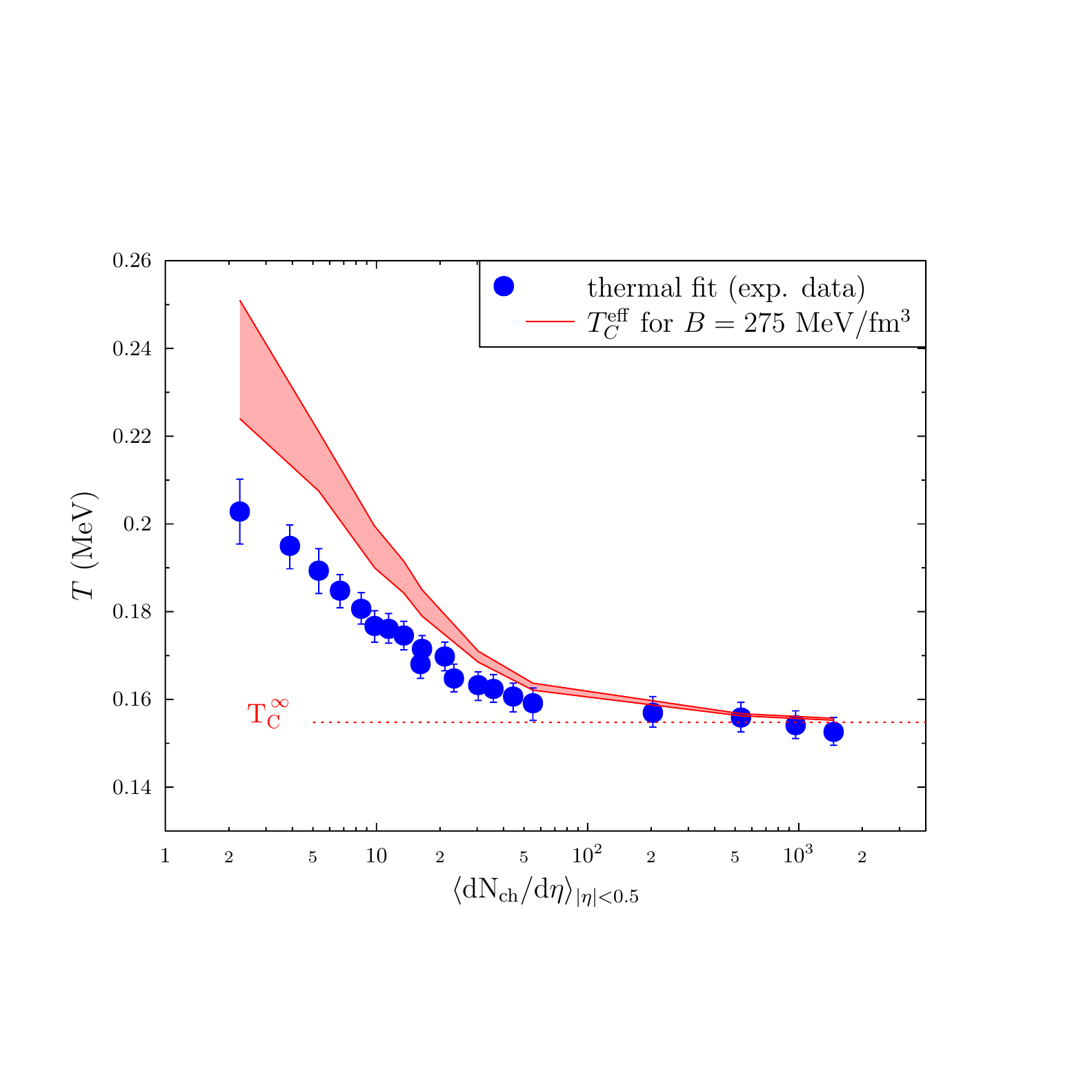}}
\vspace{-1.5cm}
\caption{The relevant critical temperature from the two-phase model as a function of charged particle multiplicity for proton-proton, proton-nucleus and nucleus-nucleus collisions at TeV energies (shown as a red band).
Also shown are the hadrochemical fits to the same experimental data within a canonical description with exact conservation of baryon number, strangeness and electrical 
charge \citep{Sharma2019}. }
\label{fig:cleymans}
\end{figure}

The hadrochemical composition of the final state of any type of high energy collision
has been successfully described in terms of thermal models, see, e.~g.
Refs.~\cite{Andronic2017,Stock2019}. In a recent
work, the charged particle multiplicity from a wide range of system sizes
has been systematically analyzed using different statistical ensembles
\citep{Sharma2019}. It turns out that the most rigourous theoretical ansatz
reviewed by the authors, the
canonical ensemble with exact conservation of strangeness, baryon number and
electrical charge, exhibits an interesting feature, namely an apparent
system size dependence of the extracted chemical freeze-out temperature.
This is shown in Fig.~\ref{fig:cleymans}, where results of Ref.~\cite{Sharma2019}
are reproduced: The freeze-out temperature corresponding to final states of
collisions with low charged particle multiplicity is significantly higher
than for high charged particle multiplicities. This is the behaviour we expect
from the schematic two-phase coexistence model, if the transition
temperature of the quark-gluon plasma created in a high energy collision, 
determines the hadrochemical freeze-out state. In Fig.~\ref{fig:cleymans},
we have plotted the effective critical temperature from the two-phase coexistence
model as a function of the charged particle multiplicity. The corresponding system
volumes are determined by the freeze-out radii from Ref.~\cite{Sharma2019}. 
The width of the resulting temperature band reflects the
error bars of the extracted freeze-out radii. Interestingly, analyzing the volume scaling of the hadrochemical fits to the experimental data, one obtains $\lambda \approx 0.5$.

\section[]{Susceptibilities of baryon number in the finite system}
In order to probe the phase structure of strongly interacting matter, event-by-event fluctuations and correlations in 
relativistic heavy-ion collisions have been proposed as a promising approach
(see, e.~g., Ref.~\cite{Luo2017}. Measured cumulants of conserved quantities can be
compared with corresponding susceptibilities from lattice QCD calculations
\citep{Bazavov2017,Guenther2018}. 
They are believed to provide rather robust signatures of the underlying thermodynamics.
The susceptibilities of the baryon number can be calculated as 
\begin{equation}\label{eq:chi}
\chi^B_i = -\frac{\partial ^i \hat{\varphi}}{\partial \hat{\mu}^i_B} \quad ,
\end{equation}
from the dimensionless density of the grand canonical potential $\hat{\varphi}=
\Phi (T,\mu_B,V) V^{-1} T^{-4}$, where $\hat{\mu}_B =\mu_B/T$ is the reduced baryochemical potential.
The second order susceptibility is proportional to the variance of the net-baryon number:
\begin{equation}
\chi^B_2(V) V T^3 = \sigma_B^2 = <(\delta N_B)^2> \quad .
\end{equation}
The ratio of the fourth to second order susceptibility is of particular
interest, since the volume and temperature terms in the definition of the 
susceptibilities (\ref{eq:chi}) cancel out, when the ratio is used. This
ratio can be measured experimentally as
\begin{equation}
\frac{\chi^B_4}{\chi^B_2} = \kappa_B\sigma_B^2 \quad,
\end{equation}
where the excess kurtosis is given by
\begin{equation}
\kappa_B = \frac{<(\delta N_B)^4>}{<(\delta N_B)^2>} - 3 \quad .
\end{equation}

\begin{figure}[b]
\hspace*{-0.7cm}
\includegraphics[width=0.59\textwidth]{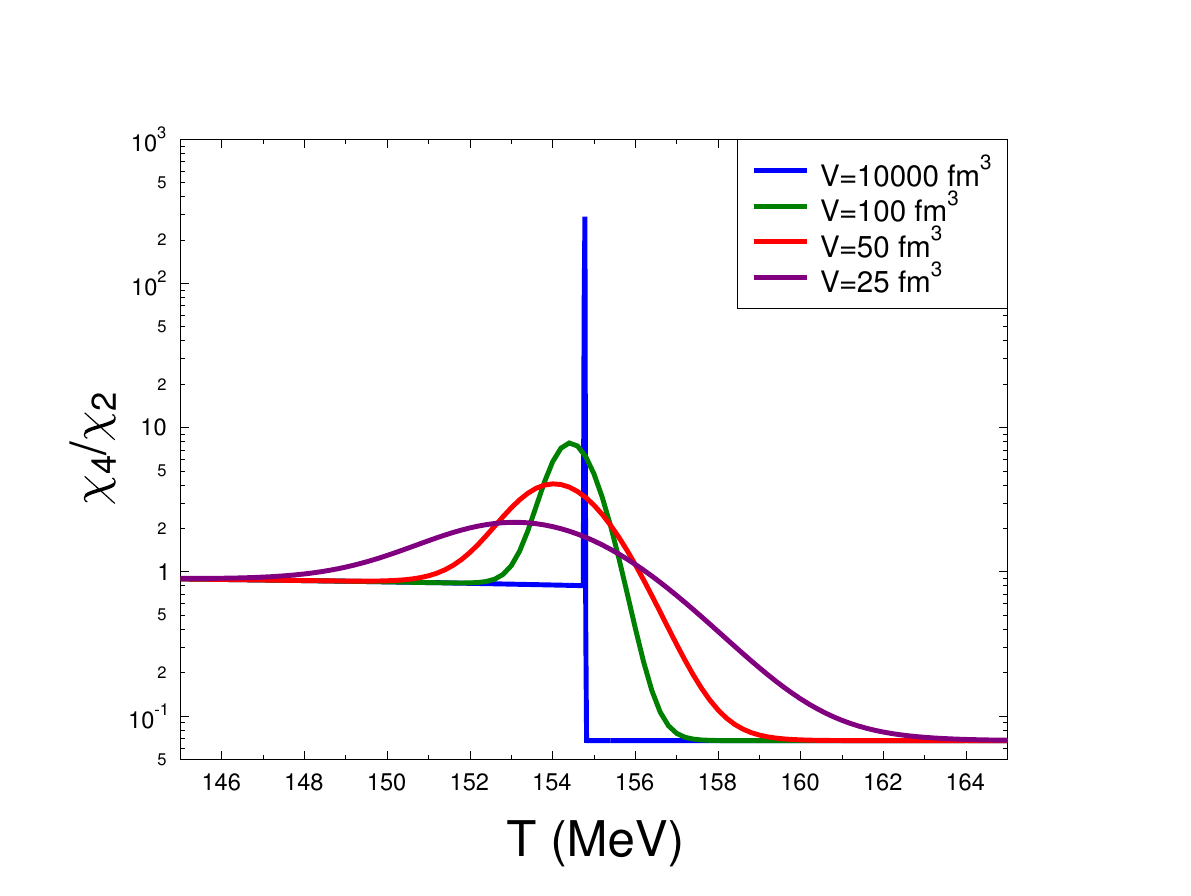}
\vspace{-0.75cm}
\caption{
Fourth to second order baryon number susceptibility ratio $\chi^B_4/\chi^B_2$ as function of
temperature at $\mu_B=0$ (Figure from Ref.~\cite{Spieles2019}). The quark-gluon phase
is not constrained to be color-neutral. } 
\label{fig:chi4chi2inf} 
\end{figure}

\begin{figure}[t]
\centering
\centerline{\includegraphics[width=250pt]{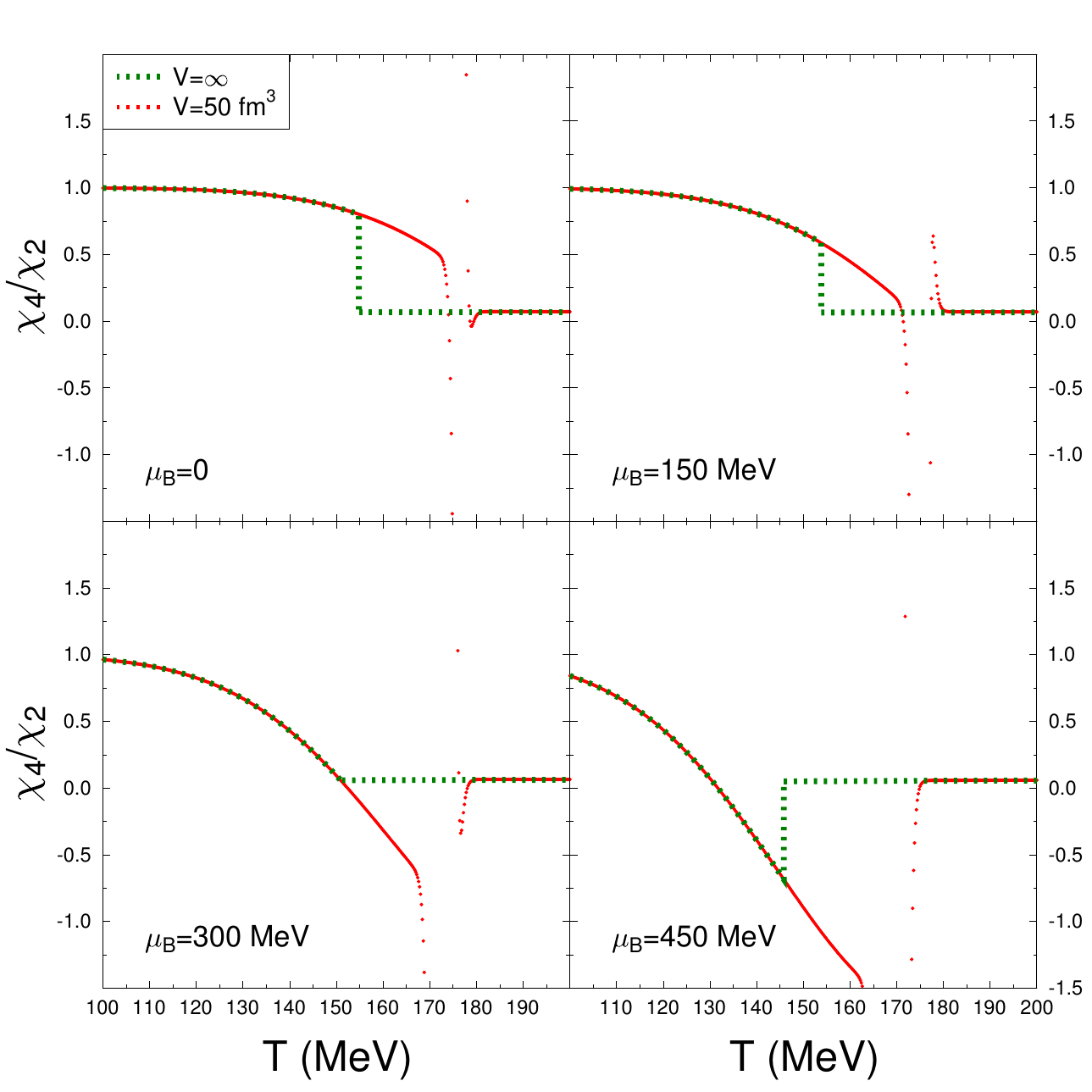}}
\caption{
Fourth to second order baryon number susceptibility ratio $\chi^B_4/\chi^B_2$ as function of
temperature at different values of baryochemical potential $\mu_B$ (Figure from
Ref.~\cite{Spieles2019}). 
The quark-gluon phase is constrained to be color-neutral. The results
for system size of $V=50~{\rm fm^3}$ (red dots) are compared with the infinite
volume limit (green dots).
} 
\label{fig:chi4chi2mu} 
\end{figure}

We begin our analysis of baryon number susceptibilities in
finite volumes by
employing an equation-of-state for the deconfined phase {\it without explicit
volume dependence} due to color-neutrality, i.e. 
we use $\phi_q(V=\infty)$ according to (\ref{eq:partitionfunc})
in (\ref{eq:phixi}). By this, we want to isolate the effect of fluctuations
of the two-phase composition in a finite system. 
Although suppressed exponentially, the presence of the
quark-gluon phase below the critical temperature $T^{\infty}_C$ and the presence of
the hadronic phase above $T^{\infty}_C$ has a finite probability.
Figure~\ref{fig:chi4chi2inf}
shows the fourth to second order baryon number susceptibility ratio $\chi^B_4/\chi^B_2$ as function of
temperature at $\mu_B=0$ for different system sizes. 
Since the two-phase coexistence model represents a physical system with a
first-order phase transition in the infinite-volume limit, we expect a
critical behaviour of thermodynamic quantities.
Indeed, for a large volume, $V=10^4\ {\rm fm^3}$, the two-phase model exhibits a strong divergence of 
the susceptibility ratio at $T_C$. This 
corresponds to extreme net-baryon number fluctuations on an
event-by-event basis in a small temperature range. 
However, as shown in Fig.~\ref{fig:chi4chi2inf}, the divergence of $\chi^B_4/\chi^B_2$ is damped for smaller volumes
by orders of magnitude. 

This is plausible since a finite system implies a maximum correlation
length. The absolute effect of
the system fluctuating between the two phases is limited by the boundaries
of the finite volume.
Note that in all cases, the value of $\chi^B_4/\chi^B_2$ approaches $1$
(expected for an ideal hadron gas) for $T\ll T_C$ and $2/(3 \pi^2)$ (expected for a gas of free, massless u/d
quarks and gluons) for $T \gg T_C$.
We conclude that in real experiments of colliding nuclei, theoretically revealing signatures
of the phase structure like "critical fluctuations" might be strongly smeared out and
suppressed due to the limited reaction sizes.

In the following, we drop the unrealistic simplification of a quark-gluon
phase without color-singlet constraint. As was discussed in
Sec.~\ref{sec:voldep}, the model exhibits a volume-dependent shift of the
effective critical temperature independent of the baryochemical potential. This effect is naturally reflected in the
baryon number susceptibilities.
Fig.~\ref{fig:chi4chi2mu}
shows the fourth to second order baryon-number susceptibility ratio $\chi^B_4/\chi^B_2$ as
a function of
temperature for $V=50~{\rm fm^3}$ for different values of  $\mu_B$,
contrasted with the respective infinite volume case.
Even  for $V=\infty$, the susceptibility ratios exhibit an
interesting feature due to the properties of the individual phases: The
hadronic resonance gas with Hagedorn-correction shows strongly reduced
values of $\chi^B_4/\chi^B_2$ with increasing baryochemical potential at
temperatures close to the phase transition. The
susceptibility ratio of the pure quark-gluon phase, on the other hand, is
virtually independent of  $\mu_B$ and $T$. 
As a consequence,  for low values of $\mu_B<300\ {\rm MeV}$, the
susceptibility ratio of the hadronic phase is significantly higher than
that of the quark-gluon phase at $T_C^{\infty}$, while the contrary is true
for $\mu_B>300\ {\rm MeV}$.
Now we compare the described characteristics of the infinite-volume
equations-of-state with the result of the two-phase coexistence model in a
finite volume, $V=50~{\rm fm^3}$. The shift of the effective critical
temperature by $\Delta T_C \approx 25\ {\rm MeV}$, represents a significant
"superheating" of the hadronic phase.\footnote{Note that the critical 
temperature $T_C^{\infty}$ itself depends on the baryochemical potential.
However, this is qualitatively irrelevant for the effect discussed here.}
This means, that the thermodynamic properties
of the hadronic phase prevail in the 
temperature range, where the suscptibility ratios of the hadronic phase have
been found to be strongly dependent on $\mu_B$, changing from values of
$\approx +1$ to $\approx -1$ within $\Delta \mu_B \approx 300\ {\rm MeV}$.
For this reason, according to the two-phase
coexistence model, the susceptibility ratios in a finite volume at low baryochemical potential
are  affected contrarily to the same system at high baryochemical potential, when
compared to infinite matter. This may point to complications in
experimental studies, where fundamental properties of strongly interacting matter are
supposed to be extracted from observable statistical observables.

\section[]{Summary}
We have demonstrated that in the schematic two-phase coexistence model of
strongly interacting matter, the shift of the
effective critical temperature as a function of system size follows a scaling law independent of
the baryochemical potential. The extracted scaling exponent $\lambda \approx 0.75$
is found to be compatible with pure SU(3) lattice gauge calculations. We find
further qualitative support for the model in recent hadrochemical analyses of charged particle multiplicities 
within a canonical formulism with conserved quantum numbers.

According to the model, critical fluctuations of baryon number close to the phase transition
should be suppressed by orders of magnitude in small systems. The model
suggests that experimentally observable susceptibility ratios $\chi^B_4/\chi^B_2$
in finite systems could be larger {\it or} smaller than in the infinite volume limit
--- depending on the baryochemical potential.

\begin{acknowledgments}
This work was supported by the Helmholtz International Center for FAIR within the framework of the LOEWE program launched by the State of Hesse.
The computational resources were provided by the Center for Scientific Computing (CSC) of the Goethe University Frankfurt.
This work has been supported by COST Action THOR (CA15213). 
\end{acknowledgments}

\bibliography{proceedings}%

\end{document}